# Optical properties of $Er^{3+}$ doped alkali-chloro phosphate glasses for optical amplifiers

K. Pradeesh[1], C.J. Oton[2], V. K. Agotiya[1], M. Raghavendra[1]

and G. Vijaya Prakash[1],♣

[1]Nanophotonics Laboratory, Department of Physics, Indian Institute of Technology Delhi, New Delhi, 110016, India
[2]Optoelectronics Research Centre, University of Southampton, Southampton, SO17 1BJ, UK

**Abstract:** A new class of Erbium doped glasses with compositions $xNa_2O - (60-x)PbCl_2 - 40P_2O_5$ (x=0, 10, 20 and 30) were fabricated and characterized for optical properties. Absorption spectra were analyzed for important Judd-Ofelt parameters from the integrated intensities of various $Er^{3+}$ glass absorption bands. Photoluminescence (PL) and its decay behavior studies were carried out for the transition $^4I_{13/2} \rightarrow {}^4I_{15/2}$. A systematic correlation between the Judd-Ofelt parameter $\Omega_2$ and the covalent nature of the glass matrix was observed, due to increased role of bridging oxygens in the glass network. The PL broadness and life times of $^4I_{13/2} \rightarrow {}^4I_{15/2}$ transition were typically in the range of 40-60nm and 2.13-2.5ms respectively. These glasses broadly showed high transparency, high refractive index, shorter life times and, most importantly, these glasses were found to be capable of being doped with larger concentrations of $Er^{3+}$ (up to 4 wt%). Increase of $Er^{3+}$ concentration resulted in the increase in PL line-widths with no significant effect of concentration quenching, indicating that these glasses are suitable for optical fibre/waveguide amplifiers.

*PACS*: 42.79; 79.60.H; 42.70.Hj

*Keywords:* Lead bearing glasses; Judd-Ofelt analysis; photoluminescence.

**Introduction:**

Incoherent broadband optical sources, particularly in telecommunication window, have been widely used in variety of applications including fiber-optic gyroscopes and optoelectronic light source for wavelength division multiplexing systems, such research resulted in semiconductor quantum dot light-emitting diodes, super-luminescent diodes, Erbium doped fibre amplifier (EDFA, C-band 1530-1565nm and L band 1565-1625nm), Erbium doped planar waveguide amplifier (EDWA), Fibre Raman amplifier, Thulium doped fibre amplifier (TDFA, S-band 1460-1530nm) etc. Though individual or co-doped rare earth doped glass fibers/waveguides are of great promise, the focus has been concentrated mostly on silicate glasses although their Amplified Spontaneous Emission (ASE) bandwidths are limited to few tens of nanometers (~40nm). Quite recently, rare-earth co-doped fluride and tellurite glasses have shown ASE with considerably extended bandwidth, where the emission broadness arises due to inter-ion energy transfer. By and large, the suitability of rare-earth doping and proper choice of glass matrix is still unclear [1-5].

In general, application and utilities of glassy materials are enormous and are governed by the factors like composition, refractive index and dopants/impurities present in the glasses. Moreover, the rare earth emission in glassy matrix is strongly dependent on crystal field effects, local environment where the ion is situated, phonon energies, refractive index and precise details about glass defects (Urbach tails) extended into the band gap. Silicate glasses, mostly being inexpensive, are preferred as host matrix but they have disadvantages such as low refractive index and strong OH band contribution. Among the other advantageous soft glassy systems, phosphate glasses have been found to be most suitable host for rare earths. For instance, phosphate glass lasers are already in the market. Significant effects have been made in the recent past in the direction of waveguide and fiber phosphate glass optical amplifier devices [6-8]. However, currently available information on phosphate matrix is still largely incomplete due to unusual structural characteristics such as high co-ordinations of the elements present and the large number of chemical elements that are being used in the compositions of these glasses. Our recent extensive studies on phosphate glasses reveal the potentiality of these glasses as broadband sources because of their wide region of transparency, moisture and thermal durability and ability in accepting large amount of rare earths as dopants [9-12].

The present paper deals with, for the first time, the absorption and emission properties of Erbium doped Alkali chloro-phosphate glass systems. Absorption spectral intensities of $Er^{3+}$ transitions in various glass compositions are analyzed by Judd-Ofelt theory. The results are examined with respect to glass composition and concentration effects, and are compared with the other reported glass systems.

**Experimental Details**:

The glass compositions of $x$ $Na_2O$ - $(60-x)$ $PbCl_2$ - $40P_2O_5$ ($x=0, 10, 20$ and $30$) (NPPx) were prepared in batches of 10g. Stoichiometric amounts of $(NH_4)_2HPO_4$, $PbCl_2$ and $NaNO_3$ (Analytical grade) were mixed with required wt% of $ErCl_3$ (Aldrich Chemical Co., USA) and were taken in a mortar and ground thoroughly using spectroscopic grade proponal. Then the mixer was taken into a silica/porcelain crucible and placed in a furnace. The temperature was raised slowly to 300 °C and was maintained for about an hour and then increased up to 600 °C , to ensure a complete decomposition of $(NH_4)_2HPO_4$ into $P_2O_5$ . Subsequently the temperature of the mixture was raised to 900

°C to get a clear melt. The homogeneity of the product was ensured by repeated stirring of the melt. The bubble free and homogeneous melt was poured on a preheated (around 250°C) brass mould and allowed to cool slowly. Good transparent bulk glasses of size of 1cm x 2cm x 0.2 cm were obtained by this process and all the glasses were annealed at 200°C for 24 hrs. The glass samples were polished to the commercial quality using a water free lubricant.

Various glass compositions with 1wt% of $Er^{3+}$ dopant were prepared for all glass compositions, mentioned above. Glasses with varying $Er^{3+}$ concentration (0.5 - 4 wt %) for one of the glass composition ,NPP20 were also fabricated, to examine the effect of concentration. However, for $Er^{3+}$ concentrations higher than 4 wt %, the glass transparency was lost. Similarly the glass transparency reduced when $Na_2O$ content was increased beyond 40, as the glass became moisture sensitive.

The absorption spectra were recorded on a Perkin-Elmer UV-VIS spectrophotometer. A 488nm wavelength of Ar-Ion laser was used as excitation source to record emission characteristics. The excitation power density was 0.5 W/cm$^2$. To measure emission and its decay curves, a cooled InGaAs detector, monochromator, mechanical chopper (12Hz), lock-in amplifier, current amplifier and digital storage oscilloscope were employed. Refractive index measurements for undoped glasses were carried out by Brewster angle setup consisting of He-Ne laser and a detector.

**Theory:**

The molar refractivity ($R_m$) was calculated by [13]

$$R_m = \frac{\eta^2 - 1}{\eta^2 + 2}\left(\frac{M}{\rho}\right) = \frac{4\pi\alpha_m N_A}{3} \qquad (1)$$

where η is the refractive index, M is the molecular weight, ρ is the density, Na is Avogadro's number and $\alpha_m$ is the polarisability of the molecule. Molar volume ($V_m$) and dielectric constant (ε) were obtained using the relations $V_m = M/\rho$ and $\varepsilon = \eta^2$ respectively. The optical band gap ($E_{opt}$) and Urbach Energies were calculated from absorption spectra of undoped glasses, using the relations discussed elsewhere [9]. The experimental oscillator strengths of various absorption spectral transitions of $Er^{3+}$ doped glasses were obtained from [14,15]

$$f_{exp} = \frac{2303mc^2}{N\pi e^2} \int \varepsilon(\bar{\nu}) d\bar{\nu} \quad (2)$$

where m and e are the electron mass and charge respectively, c is the velocity of light and $\varepsilon(\bar{\nu})$, the extinction coefficient, given by $\varepsilon(\bar{\nu}) = \left(\frac{1}{c\ell}\right)\log\left(\frac{I_0}{I}\right)$ where c is concentration of $Er^{3+}$ in moles/litre, $\ell$ is the optical path length in cm and $\log(I_0/I)$ is the absorptivity.

The oscillator strengths for the transitions were also obtained from Judd-Ofelt theory, as

$$f_{cal} = \sigma \sum T_\lambda \left\langle f^N \psi J \left| U^\lambda \right| f^N \psi' J' \right\rangle^2 \quad (3)$$

where $T_\lambda = \tau_\lambda (2J+1)^{-1}$, σ is the mean energy of the transition in $cm^{-1}$ and $T_\lambda$ are the adjustable intensity parameters. (The matrix elements are composed of $U^\lambda$). Since matrix elements of $U^\lambda$ are insensitive to ion environment, free-ion matrix elements were taken from literature [15]. The experimental oscillator strengths were fitted by least square analysis to obtain the intensity parameters $T_\lambda$. These parameters subsequently were used to calculate Judd-Ofelt parameters $\Omega_\lambda$ (λ = 2, 4 and 6) using the expression,

$$\Omega_\lambda (\text{cm}^2) = \frac{3h}{8\pi^2 mc} \frac{9\eta(2J+1)T_\lambda}{(\eta^2+2)^2} \tag{4}$$

where η is the refractive index of the medium and (2J+1) is the degeneracy of the ground level of the particular ion of interest.

**Results and Discussion:**

The Physical and optical parameters such as refractive index(η), density(ρ), molar refractive index($R_m$), molar volume ($V_m$), optical band gap($E_{opt}$) and Urbach energies (ΔE) for the glasses with various $Na_2O$ content (at the expense of $PbCl_2$), determined from the experimental data and they are shown in Table1. The refractive index values show a decreasing trend with the increase of $Na_2O$ content, except for the glass x=10. All the glasses are typically transparent above 320nm wavelength region. Estimated Optical band gap values, from the absorption spectra, for the undoped glasses are between 3.50-3.70 eV range, which are lower than the values reported for other lead based phosphate glasses[12]. Earlier observations suggest that the introduction of $Pb^{2+}$ ions to the phosphate network creates two non-bridging oxygens (NBOs), while the $Na^+$ creates single NBOs [17,18]. It is also observed that increase of $Na_2O$ content considerably shifts the absorption edge to the shorter wavelengths, predominantly due to the creation of less number of NBOs. Contrary to those observations, in the present system which has $Pb^{2+}$ ions associated with halides as glass modifier, the optical band gap shows a decreasing trend. This could be possibly due to the influence of double bonded oxygens (DBOs) associated rare earth ions linked to phosphate network, rather than NBOs[12]. On the other hand, the widths of localized states within the optical band gap (Urbach energies

($\Delta E$) were also estimated from the absorption spectra of undoped glass. In the present work these values range between 0.14 to 0.22eV, which are as close to those values reported for other phosphate glasses [9-12]. Such lower values suggest minimum defects, leading to long-range order in the present glass systems.

Spectral intensities of different energy transitions for 1wt% $Er^{3+}$ doped chlorophosphate glasses are analyzed using Eqs. (2-4). The observed and calculated oscillator strengths for various transitions to the ground state $^4I_{15/2}$ of $Er^{3+}$ ions in this glassy matrix are given in table 2. Lower RMS values suggest a good agreement between experimental and calculated oscillator strengths. It is also observed that the $Er^{3+}$ transitions lower than 350nm are not observed in these glass systems due to the cut-off region of absorption (Fig.1). In general the oscillator strengths and positions are sensitive to the environment of the rare earth sites which are occupied in the glass network [14]. These oscillator strengths are subsequently used to evaluate important intensity parameters ($\Omega_\lambda(\lambda = 2, 4$ and 6)), known as Judd-Ofelt parameters. These parameters are obtained from experimental oscillator strengths and the matrix elements using the Eqs. (2) and (3) and using least square fitting analysis for lower RMS. In general, the $\Omega_2$ values for the $Ln^{3+}$ ions in glasses are intermediate between crystalline oxides and chelating ligands [19,20]. Earlier observations suggest that both covalency and site selectivity of rare earths with non-centrosymmetric potential contributes significantly to $\Omega_2$[21]. The other values $\Omega_4$ and $\Omega_6$ are dependent on bulk properties such as viscosity and dielectric constant of the media. Hypersensitivity nature of rareearth ion transitions are charecterstic features, that are strongly depedent on covalancy and site asymmetry. Figure 4A shows the plot between the oscillator strengths of $Er^{3+}$ intense transitions of both hypersensitive as well as non-hypersensitive, versus the sum of Judd-Ofelt parameters $\Sigma\Omega_\lambda$ ($\lambda$ = 2, 4 and 6) . As seen from the figure, the hypersensitive transitions, $^4I_{15/2}\rightarrow^2H_{11/2}$, and $^4I_{15/2}\rightarrow^4G_{11/2}$ show relatively higher oscillator strengths compared to non-hypersensitive transitions $^4F_{9/2}$ and $^4S_{3/2}$, and are intermediate between ionic and highly covalent glasses [9,21]. In our present study, the observed $\Omega_2$ values are close to that of the covalent glasses (like phosphates and telurites) and on the higher side of values reported for the ionic glasses (fluorides and oxyfluorides) reported previously [9].

Electronic polarization of materials is widely regarded as one of the most influencing parameter and many physical, linear and nonlinear optical properties of materials are strongly dependent on it. Duffy, Dimitrov and Sakka correlated many independent linear optical entities to the oxide ion polarisability of single component oxides [13, 22]. This polarisability approach, predominantly gives the insight into the strong relation between covalent/ ionic nature of materials and other optical parameters, such as optical band gap. Recently we have related optical band gaps of various binary, ternary and quaternary oxide glasses to the polarisability ( of cation) in terms of 1-$R_m$/$V_m$, known as covalancy parameter or metallization parameter [9,21]. Generally, the covalancy parameter ranges from 0.3 to 0.45 for highly polarisable cation containing oxides (such as $Pb^{2+}$ and $Nb^{5+}$), while for the alkaline and alkaline-earth ( such as $Na^+$ , $Li^+$) oxides it falls in between 0.5-0.70. Although it is quite complex to correlate optical parameters of the present glassy systems with polarisability of oxide ions, as due to the presence of halides, it is interesting to see the influence of cation ($Na^+$ and/or $Pb^{2+}$-variation) polarisability on other independent parameters. Here, in our present paper, we attempted to correlate the influence of glass matrix properties on the optical properties of the dopant, that is $Er^{3+}$ ion. Since $\Omega_2$ values are known to show dependence on covalent nature of host, we made a plot between $\Omega_2$ values and metallization parameter (1-$R_m/V_m$) in figure 4B. Other phosphate glasses [23-28] are also included in the plot for comparison. Interestingly, the $\Omega_2$ values for present chlorophosphates are monotonically increasing with the metallization parameter and lie between 0.6-0.7, with comparatively high values of $\Omega_2$. Though with respect to metallization parameters they show more of alkali oxide nature, significantly higher $\Omega_2$ indicates strong covalent nature, could possibly be due to the increased role of phosphate linkage, having more double bonded oxygens (DBOs) coordinated to the rareearth ions [12].

The near infrared photoluminescence studies of $Er^{3+}$ ions for the transition $^4I_{13/2} \rightarrow {}^4I_{15/2}$ are recorded by exciting with 488nm line of $Ar^+$ laser, that is into the absorption level of $^4F_{7/2}$. The PL features are typically exhibits a narrow line shape at ~1540nm, with overall features covering 1400 to 1650nm region, and having FWHM of 40 to 45nm (Inset of Fig 1). Low intense emission at 980nm (FWHM ≤ 30nm) (not shown here) of $^4I_{11/2} \rightarrow {}^4I_{15/2}$ transition, with the relative intensity of ~ $10^{-3}$ with respect to 1540nm transition, has also been observed. The PL measurements for different glasses doped with varied $Er^{3+}$ ion concentration ranging from 0.5 to 4 wt% were performed, to observe any influence concentration on PL behavior (Fig.3). Though no peak shift has been observed, the broadness of the emission spectra did show considerably increasing trend (FWHM 43nm to 62nm) with the increased $Er^{3+}$ ion concentration.

The PL decay behavior of $^4I_{13/2} \rightarrow {}^4I_{15/2}$ transition for various glass compositions are shown in Figure 2A. All the normalized experimental data, with respect to pump energy, are fitted with the function $y(t) = A\exp(-t/\tau)^\beta$ where β parameter found to be very close to 1 and τ is the emission life time. All the decay curves in the present case are convincingly follow a single exponential stretch. The PL life times obtained from the fit are plotted (inset of Fig2A) for different $Na_2O$ content of the glasses for 1wt% $Er^{3+}$ ion concentration. These values show a moderate increase with the increase of $Na_2O$ content, except for x=10 composition, and are generally lower compared to the values reported for other phosphate glasses [29,30].

Emission performance of rareearth doped devices is dependent on the concentration of rareearth ions, in which the maximum usable concentration is restricted as the emission could quickly reach to saturation due to concentration quenching. Concentration quenching is mostly due to the non-radiative transfer of energy to closely neighboring non-excited rare earth ions through dipole-dipole coupling, resulting into decreased or no net output of light. As the concentration is increased, the rare earth ions tend to form clustering, and as a consequence the PL lifetimes of the rare earth ions is reduced. Other quenching mechanisms are up-conversion and cross-relaxation. To demonstrate the effect of $Er^{3+}$ ion concentration ( 0.5 to 4wt%) on PL decay, we have also performed experiments on one of the glasses, NPP20 (Fig 2B). Compared to the other phosphate and Lead containing germinate glasses[29,30], the present system has not shown any significant change in the life time values between concentrations 1 to 4 wt% (inset Fig2B), which is a favorable result for having minimal concentration quenching.

**Conclusions:**

In conclusion, we have fabricated a new class of Erbium doped glasses with compositions $xNa_2O$ - $(60-x)PbCl_2$ - $40P_2O_5$ (x=0, 10, 20 and 30) . The absorption spectra are analyzed for Judd-Ofelt parameters from the integrated intensities of various absorption bands. We have demonstrated the nature of hypersensitivity of some of $Er^{3+}$ absorption transitions and Judd-Ofelt parameters and attempted to correlate the effect to the covalent nature of glass network. The photoluminescence and its decay behavior studies are made for all the doped glasses. The FWHM and decay behavior show dependence on increasing alkali content and/or decrease of heavy ion content in the phosphate glass network. The PL broadness and life times of $^4I_{13/2} \rightarrow {}^4I_{15/2}$ transition were typically in the range of 40-60nm and 2.13-2.5ms respectively. By and large, these glass systems are capable of taking larger concentration of Erbium and also show increase in the width of emission with minimum concentration quenching effects. Even with the larger concentration of alkali content, these systems show good resistance to the moisture as well. Therefore, these glass systems could be the suitable hosts for the optical amplifiers from the point of view of concentration quenching effects and durability.

Table 1: Physical and optical properties of x Na$_2$O - (60-x) PbCl$_2$ - 40P$_2$O$_5$ glasses. x=0, 10, 20 and 30 (denoted as NPP0, NPP10 NPP20 and NPP30 respectively).

| Physical/optical Property | NPP0 | NPP10 | NPP20 | NPP30 |
|---|---|---|---|---|
| $E_{opt}$ (eV) | 3.55 | 3.57 | 3.50 | 3.70 |
| Urbach $\Delta E$ (eV) | 0.14 | 0.22 | 0.18 | 0.17 |
| Refractive index ($\eta$) | 1.73 | 1.57 | 1.64 | 1.59 |
| Density $\rho$ (g/cm$^3$) | 5.67 | 5.17 | 4.68 | 3.90 |
| Molecular weight M (g/mol) | 223.64 | 202.02 | 180.41 | 158.8 |
| Molar volume $V_m$ (cm$^3$/mol) | 39.44 | 39.08 | 38.55 | 40.72 |
| Molar refraction $R_m$ (cm$^3$/mol) | 15.76 | 12.78 | 13.87 | 13.65 |
| Polarisability $\alpha$ (A$^3$) | 6.25 | 5.07 | 5.50 | 5.42 |
| Judd-Ofelt parameters and PL life times of 1wt% Er doped glasses | | | | |
| $\Omega_2 \ast 10^{-20}$ (cm$^2$) | 3.36 | 4.11 | 3.66 | 3.79 |
| $\Omega_4 \ast 10^{-20}$ (cm$^2$) | 0.51 | 0.47 | 0.34 | 0.13 |
| $\Omega_6 \ast 10^{-20}$ (cm$^2$) | 1.51 | 1.16 | 1.86 | 1.21 |
| PL life times $\tau$ (ms) | 2.34 | 2.13 | 2.44 | 2.50 |

Table 2: Experimental and theoretical oscillator strengths of the spectral transitions of 1 wt% Er$^{3+}$ in NPPx glasses. x=0, 10, 20 and 30.

| Transition to From $^4I_{15/2}$ | Energy (cm$^{-1}$) | NPP0 | | NPP10 | | NPP20 | | NPP30 | |
|---|---|---|---|---|---|---|---|---|---|
| | | $f_{obs}$ x10$^{-6}$ | $f_{cal}$ x10$^{-6}$ | $f_{obs}$ x10$^{-6}$ | $f_{cal}$ x10$^{-6}$ | $f_{obs}$ x10$^{-6}$ | $f_{cal}$ x10$^{-6}$ | $f_{obs}$ x10$^{-6}$ | $f_{cal}$ x10$^{-6}$ |
| $^4I_{13/2}$ | 6447 | 1.55 | 1.65 | 1.48 | 1.49 | 1.45 | 1.80 | 1.40 | 1.11 |
| $^4I_{11/2}$ | 10065 | 0.65 | 0.76 | 0.52 | 0.71 | 0.57 | 0.85 | 0.54 | 0.57 |
| $^4I_{9/2}$ | 12446 | 0.26 | 0.14 | 0.25 | 0.12 | 0.18 | 0.09 | 0.20 | 0.42 |
| $^4F_{9/2}$ | 15298 | 1.52 | 1.62 | 1.25 | 1.44 | 1.33 | 1.61 | 0.88 | 0.93 |
| $^4S_{3/2}$ | 18229 | 0.62 | 0.66 | 0.62 | 0.62 | 0.66 | 0.76 | 0.40 | 0.47 |
| $^2H_{11/2}$ | 19065 | 5.77 | 5.70 | 5.63 | 5.97 | 5.53 | 5.62 | 4.95 | 5.30 |
| $^4F_{7/2}$ | 20339 | 2.50 | 2.26 | 2.57 | 2.08 | 3.31 | 2.49 | 1.56 | 1.54 |
| $^4F_{5/2}$ | 22009 | 0.88 | 0.80 | 0.50 | 0.75 | 0.62 | 0.92 | 0.43 | 0.57 |
| $^4F_{3/2}$ | 22388 | 0.48 | 0.46 | 0.14 | 0.43 | 0.29 | 0.53 | 0.28 | 0.33 |
| $^2G_{9/2}$ ($^4F$,$^2H$)$_{9/2}$ | 24371 | 0.75 | 0.93 | 0.49 | 0.86 | 0.77 | 1.04 | 0.48 | 0.65 |
| $^4G_{11/2}$ | 26182 | 10.00 | 10.04 | 10.7 | 10.6 | 10.00 | 9.94 | 9.60 | 9.41 |
| RMS (x10$^{-6}$) | | 0.12 | | 0.27 | | 0.33 | | 0.17 | |

**Figure Captions**

**Figure 1.** Absorption spectra of 1wt% $Er^{3+}$ doped NPP10 Glass, along with energy transitions from $^4I_{15/2}$. Inset is the $^4I_{13/2} \rightarrow {}^4I_{15/2}$ transition photoluminescence spectra of 1wt% $Er^{3+}$ doped NPP10 glass.

**Figure 2.** (A). Photoluminescence decay of 1wt% doped $Er^{3+}$ in various compositions of NPPx glasses (x=0, 10, 20 and 30).Inset is the plot of life time variation with the compositional variation (x) and (B) Photoluminescence decay of various $Er^{3+}$ concentrations ( from 0.5 to 4wt%) in NPP20 glass. Inset is the plot between PL lifetimes Vs $Er^{3+}$ concentration.

**Figure 3.** Photoluminescence spectra ($^4I_{13/2} \rightarrow {}^4I_{15/2}$ transition) of various $Er^{3+}$ concentrations in NPP20 glass. Arrow indicates the increase of concentration

**Figure 4.** (A) Oscillator strengths ( as open symbols) of various Hypersensitive ($^4G_{11/2}$ and $^2H_{11/2}$) and non-hypersensitive ($^4F_{9/2}$ and $^4S_{3/2}$) transitions of $Er^{3+}$ ions with respect to the sum of Judd-Ofelt parameters, $\Sigma\Omega_\lambda$ ($\lambda = 2, 4$ and 6). Various other glasses are also incorporated ( as filled symbols) for comparison[9,21]. Dashed lines are guide to the eye. and (B) Judd-Ofelt parameter Vs metallization parameter ($1-R_m/V_m$). Filled circles indicate present data points while the open circles are the data taken for various glass systems reported in ref.23 to 28. Dashed line indicates the trend of the data.

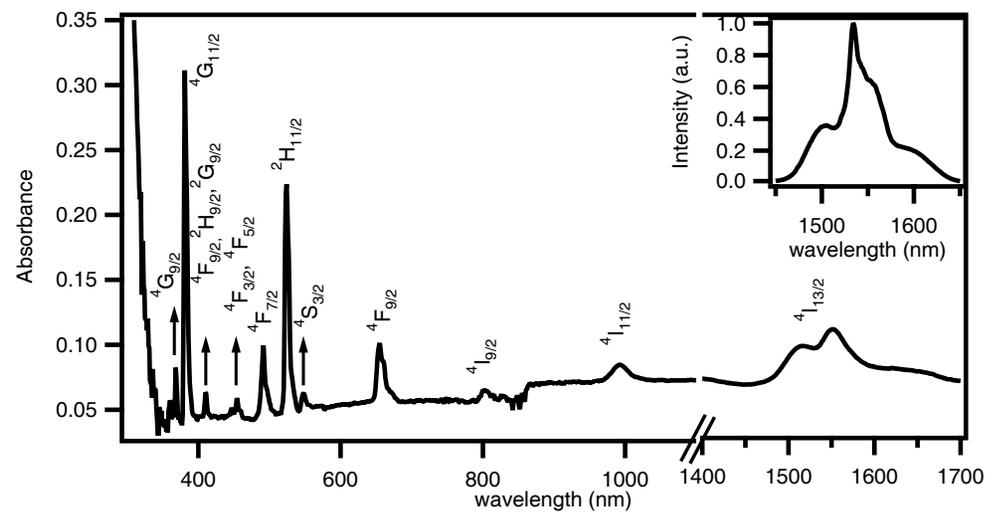

Figure 2

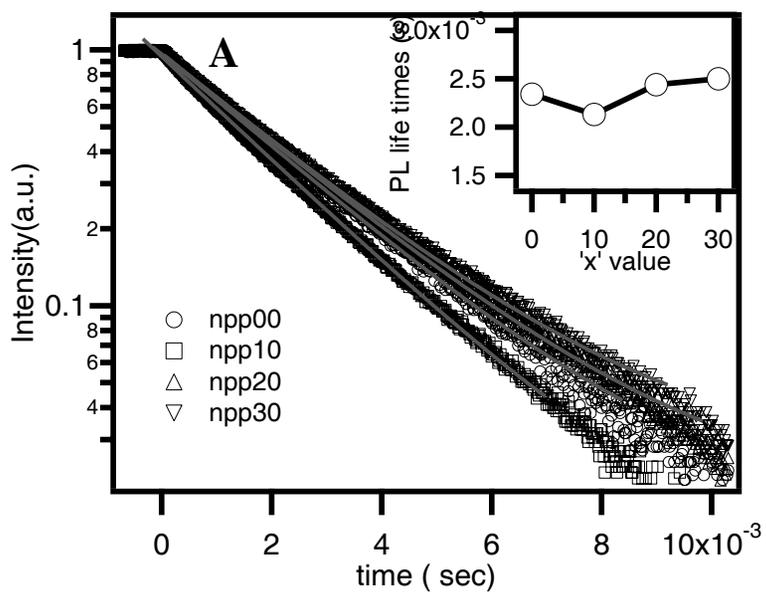
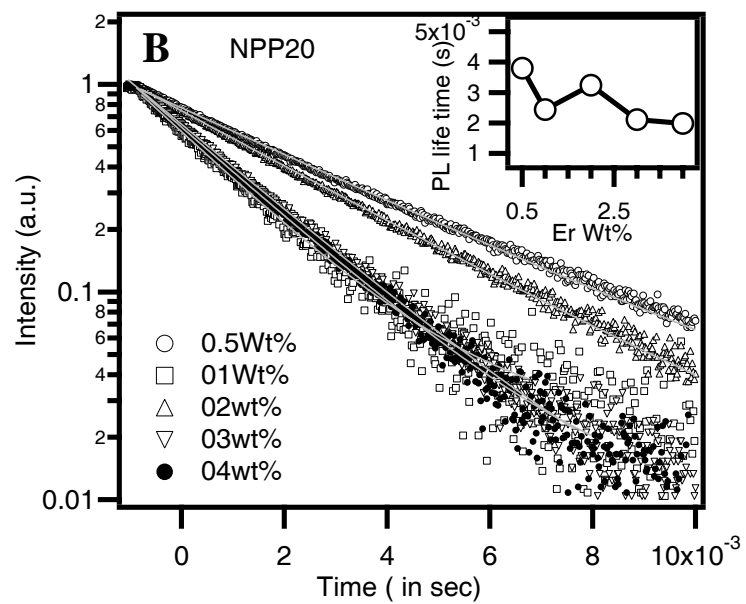

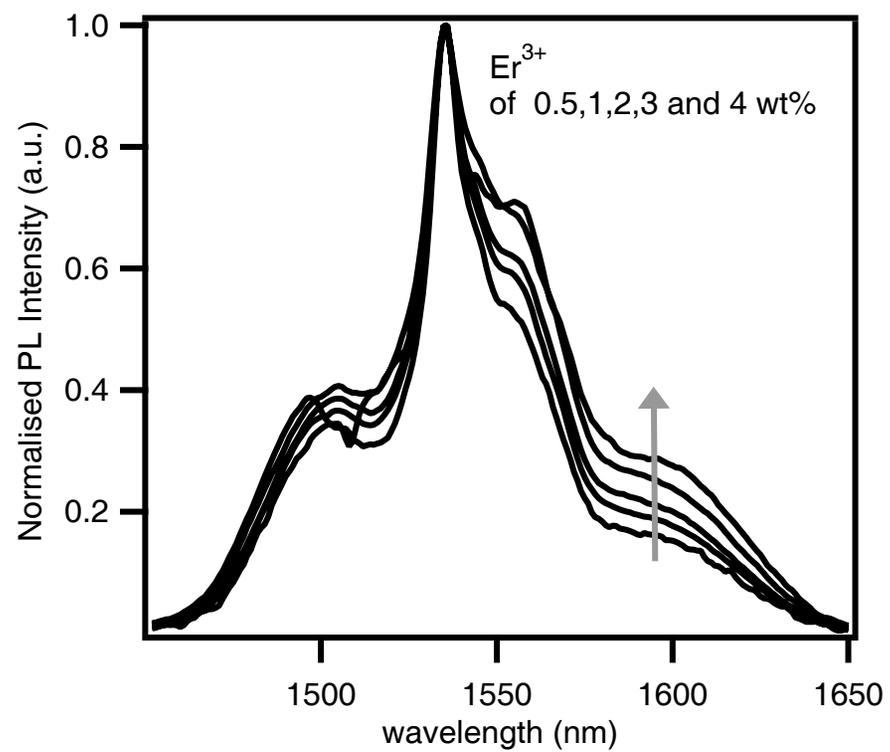

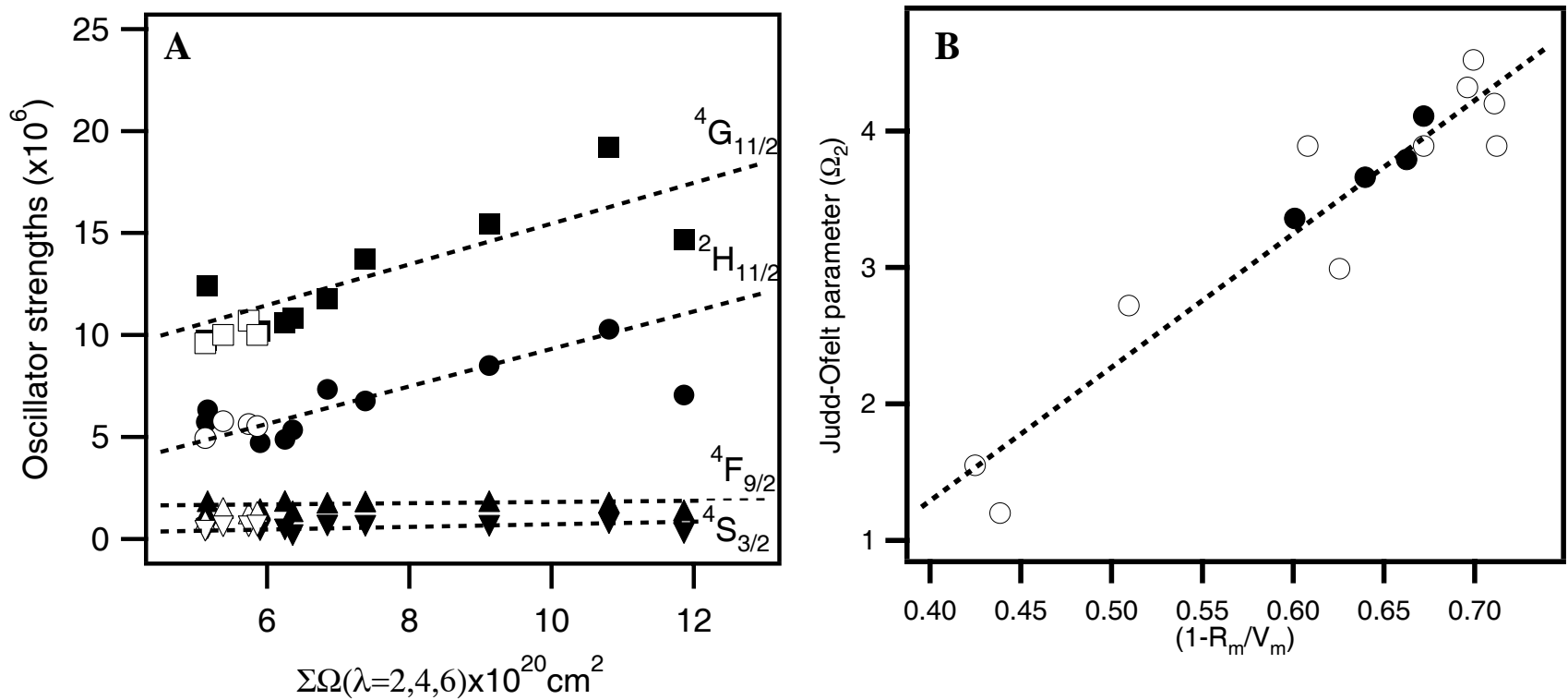